\documentclass[showpacs,floatfix,twocolumn,prb]{revtex4}

\usepackage{graphicx}

\begin{document}
\title{Theory of Core-Level Photoemission and the X-ray Edge Singularity\\Across the Mott Transition}
\date{\today}
\author{P. S. Cornaglia}
\affiliation{Centre de Physique Th\'eorique, \'Ecole Polytechnique, CNRS-UMR 7644, 91128 Palaiseau Cedex, France.}
\author{A. Georges }
\affiliation{Centre de Physique Th\'eorique, \'Ecole Polytechnique, CNRS-UMR 7644, 91128 Palaiseau Cedex, France.}

\begin{abstract}
The zero temperature core-level photoemission spectrum of a Hubbard system
is studied across the metal to Mott insulator transition using dynamical mean-field
theory and Wilson's numerical renormalization group. An asymmetric power-law divergence is
obtained in the metallic phase with
an exponent $\alpha(U,Q)-1$ which depends on the strength of both the Hubbard interaction $U$ and
the core-hole potential $Q$. For $Q\lesssim U_c/2$,
$\alpha$ decreases with increasing $U$
and vanishes at the transition ($U\to U_c$)  leading to a symmetric peak in the insulating phase.
For $Q\gtrsim U_c/2$, $\alpha$
remains finite close to the transition, but the integrated intensity of the power-law vanishes and
there is no associated peak in the insulator. The weight and position of the remaining peaks in
the spectra can be understood within a molecular orbital approach.
\end{abstract}


\maketitle
\section{Introduction}
When an incident x-ray photon ejects an electron from a core-level
in a metal, the conduction band electrons feel a local attractive
 potential due to the created hole. It was discovered by
 Anderson \cite{PhysRevLett.18.1049} that the electronic ground states
 before  and after the creation of the hole are orthogonal to each other.
 This many-body effect has dramatic consequences in x-ray photoemission
spectroscopy (XPS) experiments where an asymmetric power-law divergence is
observed.\cite{PhysRev.163.612,Nozieres1969,Doniach1970} For a 
non-interacting metal, the exponent of the power-law and the relative
 intensity of the peaks in the XPS spectra are well understood.
However, the behavior of the power-law divergence in a strongly
 interacting metal has received little theoretical attention besides
 the one-dimensional case. \cite{PhysRevLett.69.1399,Meden1998}

Recently, there have been several XPS studies of strongly correlated 
transition-metal oxides,\cite{horiba,kim:126404,taguchi,panaccione:116401} 
which addressed the changes in the core-level spectrum across the
metal to Mott insulator transition (MIT). It was observed in particular 
that the (strongly renormalized) low-energy quasiparticles present in 
the metallic phase
close to the MIT strongly modify the satellite peaks measured in
 XPS, an effect
also discussed theoretically by Kim {\it et al.} \cite{kim:126404}
In this paper, we provide a detailed theory of the core-level 
photoemission lineshape across the metal-insulator transition. 
A power-law behavior is found throughout the metallic phase up to
 the MIT where it is destroyed. Both the exponent and the intensity 
of the power-law are strongly renormalized by interactions and two 
different regimes can be identified depending
on the ratio between the valence band interaction and the core-hole
potential intensities. 

The rest of the paper is organized as follows. In Sec. \ref{sec:model} we describe the model that we use in our calculations. The numerical results for the XPS spectra across the MIT transition are presented and discussed in Sec. \ref{sec:numesp}. Section \ref{sec:molorb} contains the results of a molecular orbital approximation. Finally, we state the conclusions of our study in Sec. \ref{sec:conclus}.
\section{THE MODEL} \label{sec:model}
The simplest model to study the correlation induced MIT is the Hubbard model
\[
H = -t\sum_{\langle i,j\rangle\sigma}\left(c^\dagger_{i\sigma} c_{j\sigma}+c^\dagger_{j\sigma} c_{i\sigma}\right)%
 +U\sum_i n_{i\uparrow}n_{i\downarrow} +\epsilon_d\sum_{i\sigma}n_{i\sigma},
\]
where $c^\dagger_{i\sigma}$ creates an electron with spin $\sigma$ at site $i$ and $n_{i\sigma}=c^\dagger_{i\sigma}c_{i\sigma}$.
 A key quantity for this problem is the spectral function 
$A(\omega)=-\frac{1}{\pi}\text{Im}[G_{ii}(\omega+i0)]$ that contains
information about the valence-band single particle photoemission spectrum.
It can be obtained in the limit of large lattice coordination 
using dynamical mean-field 
theory (DMFT). \cite{georges:13}
In the DMFT framework the Hubbard model reduces to an Anderson
 impurity model coupled to a non-interacting electron bath that is calculated in
a self-consistent way. 

In the XPS experiment a photon excites a core
 electron out of the sample and the resulting 
core-hole interacts with the band electrons attractively. 
In the sudden approximation considered below, 
the self-consistent bath remains unchanged 
while the energy of the effective Anderson impurity is shifted.
The core-hole potential is taken as local and momentum independent \cite{PhysRev.163.612,Nozieres1969}
\begin{equation}
H_c = (\epsilon_h -Q\sum_\sigma c^\dagger_{0\sigma} c_{0\sigma})h^\dagger h,
\end{equation}
where $h^\dagger$($h$) creates (destroys) a core-hole at site $0$, and $\epsilon_h$ is the core-level energy.
In the sudden approximation the XPS spectrum is given by the core-hole spectral density
\begin{eqnarray}\label{eq:Ah}
 A_h(\omega)&=&\sum_{\nu_f}\delta(\omega - E_{0}^i +E_{\nu}^f + \epsilon_h) |\langle\nu_f|0_i\rangle|^2,
\end{eqnarray}
where the final states $\{|\nu_f\rangle\}$ with corresponding energies $\{E_\nu^f\}$
satisfy the sum rule $\sum_{\nu_f} |\langle\nu_f|0_i\rangle|^2 = 1$ and $|0_i\rangle$
is the initial ground state with energy $E_0^i$. The spectrum is zero for energies larger than the threshold energy $\omega_T\equiv E_0^i-E_0^f-\epsilon_h$ where $E_0^f$ is the ground state energy in the final configuration.

In what follows we will consider a Bethe lattice geometry where the free density of states is given by~\cite{georges:13}
\begin{equation}\label{eq:freedens}
D(\omega)=\frac{2}{\pi D}\sqrt{1-\left(\frac{\omega}{D}\right)^2},
\end{equation}
we will set the half-bandwidth $D\equiv 2t$ as the energy unit,
and focus on half-filling ($\epsilon_d=-\mu=U/2$).
\section{Numerical RESULTS} \label{sec:numesp}
\begin{figure}[htbp]
\includegraphics[width=8.5cm,clip=true]{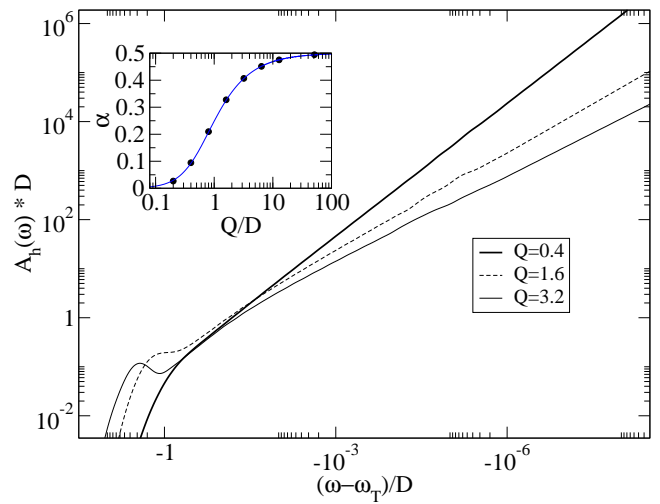}
\caption{(Color online) XPS spectra in the non-interacting case at electron-hole symmetry for different values of the core-hole potential $Q$. Inset: numerically obtained $\alpha$ (dots) and analytical expression $2/\pi^2\arctan^2(2Q)$ (solid line). }
\label{fig:nonint}
\end{figure}

In the non--interacting case ($U\to0$) the spectral function $A(\omega)$ is simply 
is given by Eq.~(\ref{eq:freedens}) and the XPS spectrum then presents a 
Doniach-\v{S}unji\'c power-law behavior at the threshold 
\cite{PhysRev.163.612,Nozieres1969,Doniach1970,RevModPhys.62.929,Oliveira1990}
\begin{equation}\label{eq:plaw_DS}
A_h^0(\omega\to \omega_T)=\frac{\pi C_0}{\Gamma(\alpha)D}\left(\frac{\omega_T-\omega}{2D}\right)^{\alpha-1},\,\, \omega <\omega_T.
\end{equation}
Here $C_0\sim 1$, $\alpha = 2 (\delta/\pi)^2= 2/\pi^2 \arctan^2(2 Q)$, 
where $\delta=\delta(\epsilon_F)$ is the phase-shift at the
 Fermi level due to scattering with the core-hole, and the factor 
$2$ is due to spin degeneracy. 
For $Q>D/2$ a bound state appears in the final configuration that is reflected in the XPS spectra as an additional peak at $\omega - \omega_T \sim -Q$. 
These analytical results have been nicely reproduced
numerically using Wilson's numerical
 renormalization group\cite{Wilson1975} (NRG) by  L\'{\i}bero 
{\it et al.}\cite{Oliveira1990}
Here we will use a different approach within the NRG scheme. 
To improve the accuracy at high energies we use a density
matrix formulation.\cite{Hofstetter2000} We have
\[
A_h(\omega)=\sum_{\nu_f,\nu_i}\delta(\omega - E_{\nu}^i +E_{\nu}^f + \epsilon_h)
\sum_{\nu_i^\prime} \langle\nu_f|\nu_i\rangle\langle\nu_i|\rho|\nu_i^\prime\rangle \langle\nu_i^\prime|\nu_f\rangle,
\]
where $\rho$ is the density matrix.
To evaluate the matrix elements $\langle\nu_f|\nu_i\rangle$ and $\langle\nu_i|\rho|\nu_i^\prime\rangle$ two simultaneous runs of NRG are performed (with and without the core-hole energy shift). The $\langle\nu_f|\nu_i\rangle$ are calculated on the first iteration and then transformed to the new basis at each NRG iteration.
The spectra are constructed following Ref. \onlinecite{PhysRevB.64.045103}.
The high numerical resolution of the NRG at low energies allows us to recover the power-law behavior with parameters that reproduce the analytical results (see Fig.~\ref{fig:nonint}). A bound state emerges in the final state for $Q>D/2$  and the associated peak in the XPS spectra can be observed at $\omega - \omega_T \sim -Q$.

For a non-zero  Hubbard interaction, the spectral function $A(\omega)$ needs
to be calculated in a self-consistent way. 
In what follows we will use the density matrix numerical renormalization 
group (NRG) scheme \cite{Hofstetter2000,anders:196801,Costi1994}
 to solve the DMFT equations~\cite{PhysRevLett.83.136,PhysRevB.64.045103} 
and to calculate the XPS spectra~\cite{Oliveira1990,Costi1996} for
different values of $U$. 
The ratio $U/D$ can be varied in some materials  across the MIT
 using chemical or mechanical pressure. The ratio $U/Q$,
 however, is expected to remain approximately 
constant for a given material. In Ref.~\onlinecite{kim:126404} the spectra in Ru 3d systems were fitted using $Q = 1.25 U$.

\begin{figure}[htbp]
\includegraphics[width=8.5cm,clip=true]{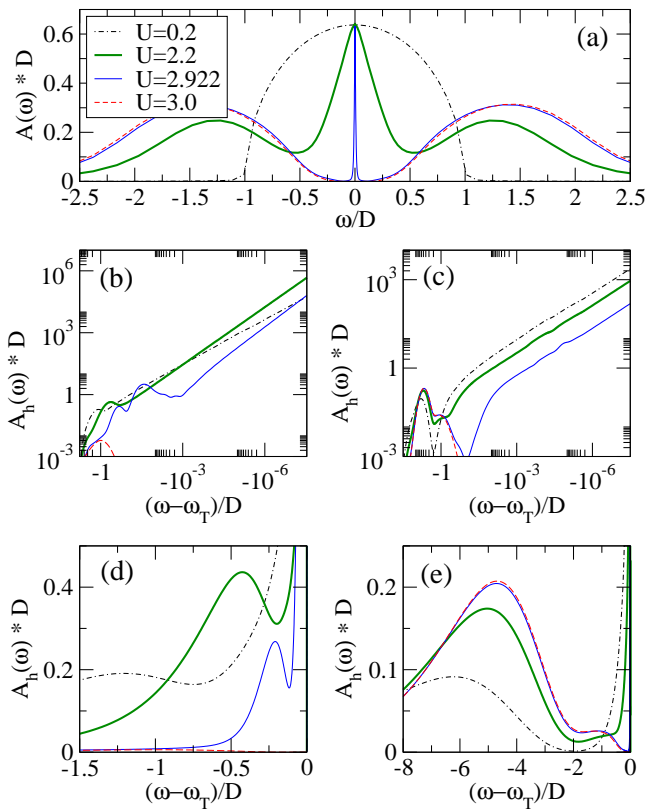}
\caption{(Color online) a) Valence spectra across the Mott transition in the electron-hole
 symmetric case. Note that for $U=3.0$ the spectrum is gapped while for $U=2.922$
 there is a narrow peak at the Fermi level. b) Core-hole spectra for a 
core-hole potential $Q=1.2$ across the MIT showing a power-law behavior
 close to the threshold in the metallic phase. c) Same as b) with $Q=6.4$.
 d) and e) Detail of the high energy peaks of b) and c), respectively.}
\label{fig:fig1}
\end{figure}

Figure~\ref{fig:fig1}a shows numerical results for the
valence band spectral function. A small interaction in the band ($U \ll D$) 
produces only a slight modification of $A(\omega)$ that close to the Fermi
level is approximately given by Eq.~(\ref{eq:freedens}). 
For intermediate values of the interaction 
($1.5 \lesssim U <U_c \simeq 2.95$) the system is
in the strongly correlated regime and the valence band
has a three three peak structure.
The central {\it quasiparticle} peak has a width 
 $\sim zD$, where $z^{-1}=1-\partial \Re[\Sigma(\omega)]/\partial\omega|_{\omega=0}$ 
and $\Sigma(\omega)$ is the self-energy. The incoherent upper and lower Hubbard bands
have a width $\sim D$ and their positions are determined by the atomic energies.
The Fermi liquid behavior is restricted to a narrow region close to the Fermi level
inside the quasiparticle peak which narrows with increasing interaction as $z \sim 0.3(U_c-U)$.
For values of $U$ larger than the critical interaction $U_c$ the system is in
 the Mott insulator phase, there is no quasiparticle peak
and the charge excitations are gapped.

Within the DMFT framework the interactions in the band produce a twofold modification
of the non-interacting x-ray problem. The atomic level coupled to the
 core-hole potential has now a Hubbard interaction $U$
 and is hybridized with a renormalized electronic bath.
To lowest order in $U$ we may neglect the changes in the valence band and
perform a Hartree-Fock approximation to treat the local 
interaction. We are left with a non-interacting
x-ray problem with a renormalized core-hole potential $Q^\prime = Q - (n-1)U/2$, 
where $n>1$ is the charge on the impurity in the presence of the core-hole potential.
In general we have $Q^\prime<Q$ and we therefore expect the bound state peak (if present) 
to be shifted toward the threshold ($\omega-\omega_T\sim -Q^\prime$) and the power-law to become more divergent
 as  $\alpha$ is reduced to $\alpha_{HF} \sim 2/\pi^2 \arctan^2(2Q^\prime)$.

In Fig.~\ref{fig:fig1}b-e we show XPS spectra for different values
of the interaction $U$ and the core-hole potential $Q$.  
We obtain a power-law behavior (straight lines) throughout the metallic phase that
disappears at the MIT.
 For small $U=0.2$ the XPS spectrum can be understood within the Hartree-Fock scheme.
This approximation, however, breaks down rapidly with increasing $U$ as
the system approaches the strongly correlated regime.
We may argue in this regime that the width of the quasiparticle peak $zD$
 acts as an effective bandwidth for the x-ray problem.~\endnote{
A related phenomenon was discussed in Ref.~[\onlinecite{Costi1996}].} 
While this allows us to understand the decrease
of the onset energy for the power-law behavior (see Figs.~\ref{fig:fig1}b and \ref{fig:fig1}c)
it does not give the complete picture as we will see below.

Close to the MIT there are some important differences in the spectra for the two values of $Q$
shown in Fig.~\ref{fig:fig1}. While the power-laws in Fig.~\ref{fig:fig1}b 
for $Q=1.2< U_c/2$ have a significant increase in the slope when approaching the MIT 
(we have $\alpha\to 0$ for $U\to U_c$), 
those for $Q=6.4$ in Fig.~\ref{fig:fig1}c have nearly parallel 
displacements with only a very small increase in the slope 
throughout the metallic phase. 
For $Q=1.2$ the high energy peaks shift toward the edge 
with increasing $U$ and disappear at the MIT leaving a delta-function 
peak at the threshold in the insulator. For $Q=6.4>U_c/2$ the main difference
in the XPS spectra just below ($U=2.922$) and above ($U=3.0$) 
the MIT is the presence of the power-law in the metallic phase. 
The high energy peaks are essentially unchanged
and the integrated intensity of the power-law vanishes at the MIT with $\alpha\neq 0$.

Some insight on the origin of these two distinct behaviors can be obtained
by analyzing the atomic limit in the insulator. For values of $Q$ such that
$Q<U_c/2$, we have $Q<U/2$ throughout the insulating phase, which
means that the  core-hole potential is not strong enough
(in the electron-hole symmetric case) to overcome the local
repulsion and change the charge locally in the atomic level. 
The initial and final ground states will be identical and we
 therefore expect a peak at the threshold. 
In the opposite situation, $Q>U_c/2$, there will be a region close to
the MIT in which the initial and final states have a different
 occupation of the atomic level.  
For these values of $Q$, the initial and final ground states are orthogonal
and no peak is expected at the threshold.

\begin{figure}[htbp]
\includegraphics[width=8.5cm,clip=true]{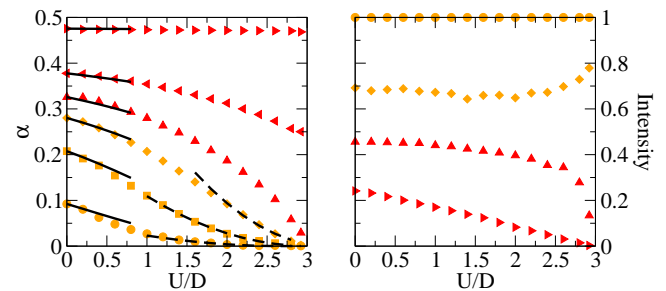}
\caption{(Color online) Left panel:
Exponent $\alpha$ as a function of $U$, for different values of
$Q=12.8,\,2.4,\,1.6,\,1.2,\,0.8,$ and $0.4$ (from top to bottom).
Results from NRG (symbols), Hartree-Fock results (solid lines), and parabolic fits $f(U)=A Q^2(U_c-U)^2$ with $A\sim0.04$ (dashed lines).
Right panel: integrated intensity of the power-law peak for $Q=12.8,\,1.6,\,1.2,$ and $0.4$
(from bottom to top).
The figure illustrates that the peak at threshold disappears at the MIT in
two different manners, for
$Q\gtrsim U_c/2$ and $Q\lesssim U_c/2$ (see text).
}\label{fig:p-shiftU}
\end{figure}

The results for the exponent and the integrated intensity of the power-law are
summarized in Fig.~\ref{fig:p-shiftU}. 
The exponent $\alpha$ increases with $Q$ and decreases with
increasing interaction. For small $U\ll Q, D$ the main effect
 of the interaction is a compensation (screening) of the core-hole
 potential that can be explained within a Hartree-Fock approach 
(solid lines in Fig.~\ref{fig:p-shiftU}). For large $Q\gg U_c$, $\alpha$ 
is close to its maximum value $1/2$ and the interaction produces only 
a small reduction throughout the metallic phase. The intensity of the
 peak at the threshold, however, vanishes continuously at the MIT. 
For small $Q<U_c/2$ the opposite behavior is observed with $\alpha$ 
vanishing at the transition while the integrated intensity of the peak remains finite.

The behavior of $\alpha$ can be understood through a Fermi liquid analysis.
The Friedel sum rule gives an exact zero-temperature relation between 
the scattering phase-shift and the {\it total} charge
 $\Delta N$ ($d$-orbital plus conduction band) displaced when the 
core-hole potential is turned on, \cite{Friedel1952,PhysRev.150.516}
namely: $\delta = \pi \Delta N/2$.
In the insulating phase, the valence band has a gap and the core-hole
 potential $Q$
will not produce a displacement of charge ($\Delta N\equiv 0$) unless $Q$ is
large enough to overcame the local repulsion (i.e. $Q>U/2$). This reflects the incompressibility of the
insulator. In contrast, in the metallic phase,
charge will gather around the hole to screen it even for $Q\ll U$.
The amount of displaced charge reflects the vanishing compressibility of the
metal as the MIT is approached, so that we expect at small $Q$:
$\Delta N \propto z Q$. Since Fermi-liquid analysis suggests that the Nozi{\`e}res and de Dominicis' value
for the exponent [$\alpha=2(\delta/\pi)^2$] is valid throughout the metallic phase even in
the presence of interactions, this leads to a parabolic behavior
$\alpha\propto 2\,(Q z/2)^2 \sim 0.05\, Q^2(U_c-U)^2$,
[where we used $z\sim 0.3 (U_c-U)$].
This expression fits our data remarkably well, close to $U_c$ and for $Q<U_c/2$
(see Fig.~\ref{fig:p-shiftU}).

\begin{figure}[htbp]
\includegraphics[width=8.5cm,clip=true]{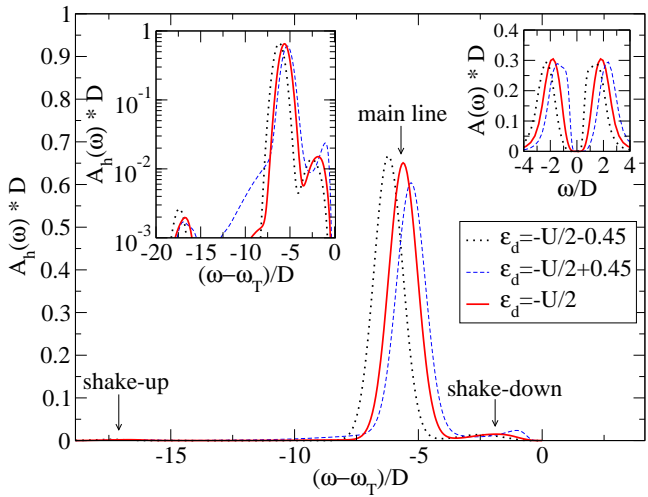}
\caption{(Color online)
Core-hole photoemission spectra in the insulating phase ($U=3.8$)
for fixed $Q=2U$ and three values of $\epsilon_d$. Left inset: in a
logarithmic scale three peaks can be identified. Right inset:
Hubbard bands in the valence spectra.}
\label{fig:insulating}
\end{figure}

\section{Molecular Orbital approximation}\label{sec:molorb}
\begin{figure}[htbp]
\includegraphics[width=8.5cm,clip=true]{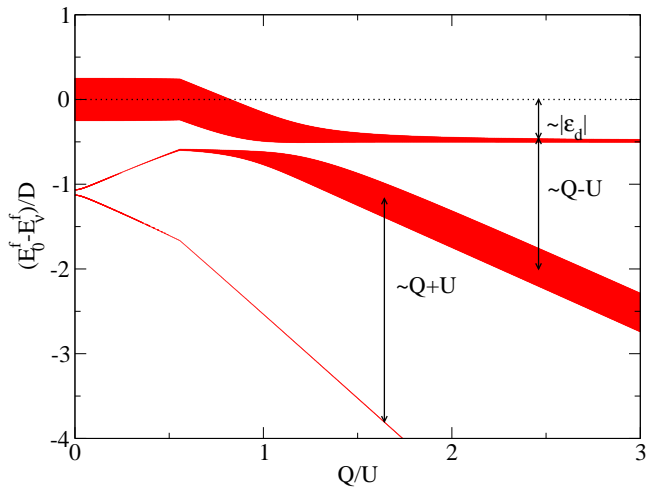}
\caption{(Color online) Molecular orbital results for the position of the XPS peaks in the insulating phase.  The thickness of the lines is proportional to the peak intensity. Parameters are $U=3.8D$, $\epsilon_d=-U/2$, and $\epsilon_h=0$. }
\label{fig:intensities}
\end{figure}

The intensity and position of the peaks in the insulating phase can be understood
 in the molecular orbital approximation.
The effective bath in DMFT is modeled 
with a single site having a Hubbard interaction $U$ and energy $\epsilon_d$.
 The resulting two site Hubbard model can be solved analytically. In the absence 
of the core-hole potential, the ground state is in a subspace with 2 electrons.
 When the core-hole potential is small ($Q\lesssim U/2$) the ground state is in
 the same charge sector and is essentially unchanged.
 Therefore, the XPS spectrum has a delta-function peak at the threshold which 
carries almost all the spectral weight (see Fig.~\ref{fig:intensities}).
For large $Q\gtrsim U/2$ the final ground state is in a subspace
with three electrons and is therefore orthogonal to the initial
ground state ($\langle 0_f|0_i\rangle$=0). 
In this regime three peaks can be identified
in the XPS spectra and there is no peak at the threshold.
The main peak has a weight $1-\frac{3}{Q^2} -\frac{1}{U^2}+\cdots$ and
 there are two satellites at a distance $\sim U+Q$ (shake-up) 
and $\sim Q-U$ (shake-down). 
The shake-down peak is at a distance $\epsilon_d$ from the threshold and has a larger intensity ($\frac{1}{2U^2} +\frac{1}{UQ}+\frac{3}{2Q^2}+\cdots$) than the shake-up peak  ($\frac{1}{2U^2} -\frac{1}{UQ}+\frac{3}{2Q^2}+\cdots$).

The molecular results are in excellent agreement with the numerical
 results in the insulating phase (see Fig.~\ref{fig:insulating}). 
The shake-up (-down) peak is associated to the creation of an electron 
(hole) in the upper (lower) Hubbard band and has an intrinsic width
of order $D$. For a finite hole lifetime (finite temperature) the 
spectrum will have an additional Lorentzian (Gaussian) 
broadening~\endnote{The numerical procedure employed gives an 
additional width to the high energy peaks~[\onlinecite{PhysRevB.64.045103}].}.  
We note that, although it is likely to be difficult in practice, 
a measurement of the {\it three} peaks provides in principle a direct 
estimate of $U$. We stress that a measurement of the shake-down peak 
alone is not enough for such an estimation.
As shown in Fig.~\ref{fig:insulating}, a change in
 $\epsilon_d$ only produces a small redistribution of the 
spectral intensity and a global shift of the peaks relative to the
 threshold. The molecular orbital results also give a good estimation
for the position of the peaks away from the threshold in the metallic 
phase (see Fig.~\ref{fig:fig1}e).

\begin{figure}[htbp]
\includegraphics[height=8.5cm,angle=-90,clip=true]{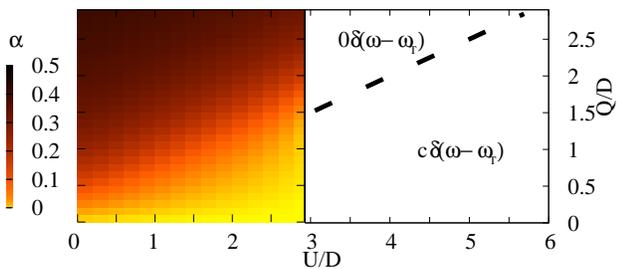}
\caption{(Color online) Behavior of the XPS spectra at the edge as a function 
of $U$ and $Q$. Left panel: coefficient $\alpha$ of the power-law
 divergence in the metal. Right panel: below the line $Q\sim U/2$ 
the spectra present a delta-function peak in the insulator.} 
\label{fig:pdiag}
\end{figure}

\section{Summary and Conclusions}\label{sec:conclus}

In summary, we have studied theoretically the behavior of the core-level photoemission spectra
across the correlation-driven MIT.
Away from the photoemission threshold and far from the MIT, both the position and relative intensity of the
peaks are well described by a molecular orbital approach. Close to the threshold or
to the MIT, more sophisticated techniques (such as DMFT and NRG) are necessary to describe the spectra.
The changes in the XPS spectra across the MIT may be used to detect the transition.
A symmetric peak (or no peak) at the edge implies an insulating phase, while an asymmetric peak or a power-law
corresponds to a metal.
For large $Q>U/2$ there are three peaks in the XPS spectra. While the shake-up peak is usually weak and
will probably be difficult to detect, its observation in conjunction with the
shake-down peak allows in principle for a direct estimation of $U$.

The most interesting results concern the behavior of the XPS 
spectra at the threshold as summarized in Fig.~\ref{fig:pdiag}.
In the metallic phase there is a power-law behavior
with an exponent $\alpha-1$, where $\alpha$ depends on both the
local interactions on the band $U$ and the intensity
 of the core-hole potential $Q$.
The power-law regime is restricted to an energy range
$\sim zD$ close to the threshold and disappears at the MIT. For $Q\lesssim U_c/2$ the exponent 
$\alpha$ depends strongly on $U$ and vanishes as $\sim z^2$ at the transition.
The integrated intensity, however, remains finite giving rise to a delta-function peak in the insulator.
In contrast, for $Q\gtrsim U_c/2$, $\alpha$ is finite at the transition but the
integrated intensity of the peak vanishes and there is no delta-function peak 
in the insulator. 
Future high-resolution experiments might be able to test our theoretical
predictions for the edge-singularity in a correlated metal
close to the MIT.

\acknowledgments
We are thankful to M.~Altarelli, G.~Kotliar, G.~Panaccione, A.~Poteryaev, and G.~Sawatzky for useful discussions.


\begin{thebibliography}{22}
\expandafter\ifx\csname natexlab\endcsname\relax\def\natexlab#1{#1}\fi
\expandafter\ifx\csname bibnamefont\endcsname\relax
  \def\bibnamefont#1{#1}\fi
\expandafter\ifx\csname bibfnamefont\endcsname\relax
  \def\bibfnamefont#1{#1}\fi
\expandafter\ifx\csname citenamefont\endcsname\relax
  \def\citenamefont#1{#1}\fi
\expandafter\ifx\csname url\endcsname\relax
  \def\url#1{\texttt{#1}}\fi
\expandafter\ifx\csname urlprefix\endcsname\relax\def\urlprefix{URL }\fi
\providecommand{\bibinfo}[2]{#2}
\providecommand{\eprint}[2][]{\url{#2}}

\bibitem[{\citenamefont{Anderson}(1967)}]{PhysRevLett.18.1049}
\bibinfo{author}{\bibfnamefont{P.~W.} \bibnamefont{Anderson}},
  \bibinfo{journal}{Phys. Rev. Lett.} \textbf{\bibinfo{volume}{18}},
  \bibinfo{pages}{1049} (\bibinfo{year}{1967}).

\bibitem[{\citenamefont{Mahan}(1967)}]{PhysRev.163.612}
\bibinfo{author}{\bibfnamefont{G.~D.} \bibnamefont{Mahan}},
  \bibinfo{journal}{Phys. Rev.} \textbf{\bibinfo{volume}{163}},
  \bibinfo{pages}{612} (\bibinfo{year}{1967}).

\bibitem[{\citenamefont{Nozi{\`e}res and De~Dominicis}(1969)}]{Nozieres1969}
\bibinfo{author}{\bibfnamefont{P.}~\bibnamefont{Nozi{\`e}res}}
  \bibnamefont{and} \bibinfo{author}{\bibfnamefont{C.~T.}
  \bibnamefont{De~Dominicis}}, \bibinfo{journal}{Phys. Rev.}
  \textbf{\bibinfo{volume}{178}}, \bibinfo{pages}{1097} (\bibinfo{year}{1969}).

\bibitem[{\citenamefont{Doniach and Sunji{\'c}}(1970)}]{Doniach1970}
\bibinfo{author}{\bibfnamefont{S.}~\bibnamefont{Doniach}} \bibnamefont{and}
  \bibinfo{author}{\bibfnamefont{M.}~\bibnamefont{Sunji{\'c}}},
  \bibinfo{journal}{J. Phys. C} \textbf{\bibinfo{volume}{3}},
  \bibinfo{pages}{285} (\bibinfo{year}{1970}).

\bibitem[{\citenamefont{Lee and Chen}(1992)}]{PhysRevLett.69.1399}
\bibinfo{author}{\bibfnamefont{D.~K.~K.} \bibnamefont{Lee}} \bibnamefont{and}
  \bibinfo{author}{\bibfnamefont{Y.}~\bibnamefont{Chen}},
  \bibinfo{journal}{Phys. Rev. Lett.} \textbf{\bibinfo{volume}{69}},
  \bibinfo{pages}{1399} (\bibinfo{year}{1992}).

\bibitem[{\citenamefont{Meden et~al.}(1998)\citenamefont{Meden, Schmitteckert,
  and Shannon}}]{Meden1998}
\bibinfo{author}{\bibfnamefont{V.}~\bibnamefont{Meden}},
  \bibinfo{author}{\bibfnamefont{P.}~\bibnamefont{Schmitteckert}},
  \bibnamefont{and} \bibinfo{author}{\bibfnamefont{N.}~\bibnamefont{Shannon}},
  \bibinfo{journal}{Phys. Rev. B} \textbf{\bibinfo{volume}{57}},
  \bibinfo{pages}{8878} (\bibinfo{year}{1998}).

\bibitem[{\citenamefont{Horiba et~al.}(2004)\citenamefont{Horiba, Taguchi,
  Chainani, Takata, Ikenaga, Miwa, Nishino, Tamasaku, Awaji, Takeuchi
  et~al.}}]{horiba}
\bibinfo{author}{\bibfnamefont{K.}~\bibnamefont{Horiba}},
  \bibinfo{author}{\bibfnamefont{M.}~\bibnamefont{Taguchi}},
  \bibinfo{author}{\bibfnamefont{A.}~\bibnamefont{Chainani}},
  \bibinfo{author}{\bibfnamefont{Y.}~\bibnamefont{Takata}},
  \bibinfo{author}{\bibfnamefont{E.}~\bibnamefont{Ikenaga}},
  \bibinfo{author}{\bibfnamefont{D.}~\bibnamefont{Miwa}},
  \bibinfo{author}{\bibfnamefont{Y.}~\bibnamefont{Nishino}},
  \bibinfo{author}{\bibfnamefont{K.}~\bibnamefont{Tamasaku}},
  \bibinfo{author}{\bibfnamefont{M.}~\bibnamefont{Awaji}},
  \bibinfo{author}{\bibfnamefont{A.}~\bibnamefont{Takeuchi}},
  \bibnamefont{et~al.}, \bibinfo{journal}{Phys. Rev. Lett.}
  \textbf{\bibinfo{volume}{93}}, \bibinfo{eid}{236401} (\bibinfo{year}{2004}).

\bibitem[{\citenamefont{Kim et~al.}(2004)\citenamefont{Kim, Noh, Kim, and
  Oh}}]{kim:126404}
\bibinfo{author}{\bibfnamefont{H.-D.} \bibnamefont{Kim}},
  \bibinfo{author}{\bibfnamefont{H.-J.} \bibnamefont{Noh}},
  \bibinfo{author}{\bibfnamefont{K.~H.} \bibnamefont{Kim}}, \bibnamefont{and}
  \bibinfo{author}{\bibfnamefont{S.-J.} \bibnamefont{Oh}},
  \bibinfo{journal}{Phys. Rev. Lett.} \textbf{\bibinfo{volume}{93}},
  \bibinfo{pages}{126404} (\bibinfo{year}{2004}).

\bibitem[{\citenamefont{Taguchi et~al.}(2005)\citenamefont{Taguchi, Chainani,
  Kamakura, Horiba, Takata, Yabashi, Tamasaku, Nishino, Miwa, Ishikawa
  et~al.}}]{taguchi}
\bibinfo{author}{\bibfnamefont{M.}~\bibnamefont{Taguchi}},
  \bibinfo{author}{\bibfnamefont{A.}~\bibnamefont{Chainani}},
  \bibinfo{author}{\bibfnamefont{N.}~\bibnamefont{Kamakura}},
  \bibinfo{author}{\bibfnamefont{K.}~\bibnamefont{Horiba}},
  \bibinfo{author}{\bibfnamefont{Y.}~\bibnamefont{Takata}},
  \bibinfo{author}{\bibfnamefont{M.}~\bibnamefont{Yabashi}},
  \bibinfo{author}{\bibfnamefont{K.}~\bibnamefont{Tamasaku}},
  \bibinfo{author}{\bibfnamefont{Y.}~\bibnamefont{Nishino}},
  \bibinfo{author}{\bibfnamefont{D.}~\bibnamefont{Miwa}},
  \bibinfo{author}{\bibfnamefont{T.}~\bibnamefont{Ishikawa}},
  \bibnamefont{et~al.}, \bibinfo{journal}{Phys. Rev. B}
  \textbf{\bibinfo{volume}{71}}, \bibinfo{eid}{155102} (\bibinfo{year}{2005}).

\bibitem[{\citenamefont{Panaccione et~al.}(2006)\citenamefont{Panaccione,
  Altarelli, Fondacaro, Georges, Huotari, Lacovig, Lichtenstein, Metcalf,
  Monaco, Offi et~al.}}]{panaccione:116401}
\bibinfo{author}{\bibfnamefont{G.}~\bibnamefont{Panaccione}},
  \bibinfo{author}{\bibfnamefont{M.}~\bibnamefont{Altarelli}},
  \bibinfo{author}{\bibfnamefont{A.}~\bibnamefont{Fondacaro}},
  \bibinfo{author}{\bibfnamefont{A.}~\bibnamefont{Georges}},
  \bibinfo{author}{\bibfnamefont{S.}~\bibnamefont{Huotari}},
  \bibinfo{author}{\bibfnamefont{P.}~\bibnamefont{Lacovig}},
  \bibinfo{author}{\bibfnamefont{A.}~\bibnamefont{Lichtenstein}},
  \bibinfo{author}{\bibfnamefont{P.}~\bibnamefont{Metcalf}},
  \bibinfo{author}{\bibfnamefont{G.}~\bibnamefont{Monaco}},
  \bibinfo{author}{\bibfnamefont{F.}~\bibnamefont{Offi}}, \bibnamefont{et~al.},
  \bibinfo{journal}{Phys. Rev. Lett.} \textbf{\bibinfo{volume}{97}},
  \bibinfo{pages}{116401} (\bibinfo{year}{2006}).

\bibitem[{\citenamefont{Georges et~al.}(1996)\citenamefont{Georges, Kotliar,
  Krauth, and Rozenberg}}]{georges:13}
\bibinfo{author}{\bibfnamefont{A.}~\bibnamefont{Georges}},
  \bibinfo{author}{\bibfnamefont{G.}~\bibnamefont{Kotliar}},
  \bibinfo{author}{\bibfnamefont{W.}~\bibnamefont{Krauth}}, \bibnamefont{and}
  \bibinfo{author}{\bibfnamefont{M.~J.} \bibnamefont{Rozenberg}},
  \bibinfo{journal}{Rev. Mod. Phys.} \textbf{\bibinfo{volume}{68}},
  \bibinfo{pages}{13} (\bibinfo{year}{1996}).

\bibitem[{\citenamefont{Ohtaka and Tanabe}(1990)}]{RevModPhys.62.929}
\bibinfo{author}{\bibfnamefont{K.}~\bibnamefont{Ohtaka}} \bibnamefont{and}
  \bibinfo{author}{\bibfnamefont{Y.}~\bibnamefont{Tanabe}},
  \bibinfo{journal}{Rev. Mod. Phys.} \textbf{\bibinfo{volume}{62}},
  \bibinfo{pages}{929} (\bibinfo{year}{1990}).

\bibitem[{\citenamefont{L{\'{\i}}bero and Oliveira}(1990)}]{Oliveira1990}
\bibinfo{author}{\bibfnamefont{V.~L.} \bibnamefont{L{\'{\i}}bero}}
  \bibnamefont{and} \bibinfo{author}{\bibfnamefont{L.~N.}
  \bibnamefont{Oliveira}}, \bibinfo{journal}{Phys. Rev. B}
  \textbf{\bibinfo{volume}{42}}, \bibinfo{pages}{3167} (\bibinfo{year}{1990}).

\bibitem[{\citenamefont{Wilson}(1975)}]{Wilson1975}
\bibinfo{author}{\bibfnamefont{K.~G.} \bibnamefont{Wilson}},
  \bibinfo{journal}{Rev. Mod. Phys.} \textbf{\bibinfo{volume}{47}},
  \bibinfo{pages}{773} (\bibinfo{year}{1975}).

\bibitem[{\citenamefont{Hofstetter}(2000)}]{Hofstetter2000}
\bibinfo{author}{\bibfnamefont{W.}~\bibnamefont{Hofstetter}},
  \bibinfo{journal}{Phys. Rev. Lett.} \textbf{\bibinfo{volume}{85}},
  \bibinfo{pages}{1508} (\bibinfo{year}{2000}).

\bibitem[{\citenamefont{Bulla et~al.}(2001)\citenamefont{Bulla, Costi, and
  Vollhardt}}]{PhysRevB.64.045103}
\bibinfo{author}{\bibfnamefont{R.}~\bibnamefont{Bulla}},
  \bibinfo{author}{\bibfnamefont{T.~A.} \bibnamefont{Costi}}, \bibnamefont{and}
  \bibinfo{author}{\bibfnamefont{D.}~\bibnamefont{Vollhardt}},
  \bibinfo{journal}{Phys. Rev. B} \textbf{\bibinfo{volume}{64}},
  \bibinfo{pages}{045103} (\bibinfo{year}{2001}).

\bibitem[{\citenamefont{Anders and Schiller}(2005)}]{anders:196801}
\bibinfo{author}{\bibfnamefont{F.~B.} \bibnamefont{Anders}} \bibnamefont{and}
  \bibinfo{author}{\bibfnamefont{A.}~\bibnamefont{Schiller}},
  \bibinfo{journal}{Phys. Rev. Lett.} \textbf{\bibinfo{volume}{95}},
  \bibinfo{eid}{196801} (\bibinfo{year}{2005}).

\bibitem[{\citenamefont{Costi et~al.}(1994)\citenamefont{Costi, Hewson, and
  Zlati{\'c}}}]{Costi1994}
\bibinfo{author}{\bibfnamefont{T.~A.} \bibnamefont{Costi}},
  \bibinfo{author}{\bibfnamefont{A.~C.} \bibnamefont{Hewson}},
  \bibnamefont{and}
  \bibinfo{author}{\bibfnamefont{V.}~\bibnamefont{Zlati{\'c}}},
  \bibinfo{journal}{J. Phys.:Condens. Matter} \textbf{\bibinfo{volume}{6}},
  \bibinfo{pages}{2519} (\bibinfo{year}{1994}).

\bibitem[{\citenamefont{Bulla}(1999)}]{PhysRevLett.83.136}
\bibinfo{author}{\bibfnamefont{R.}~\bibnamefont{Bulla}},
  \bibinfo{journal}{Phys. Rev. Lett.} \textbf{\bibinfo{volume}{83}},
  \bibinfo{pages}{136} (\bibinfo{year}{1999}).

\bibitem[{\citenamefont{Costi et~al.}(1996)\citenamefont{Costi, Kroha, and
  W\"olfle}}]{Costi1996}
\bibinfo{author}{\bibfnamefont{T.~A.} \bibnamefont{Costi}},
  \bibinfo{author}{\bibfnamefont{J.}~\bibnamefont{Kroha}}, \bibnamefont{and}
  \bibinfo{author}{\bibfnamefont{P.}~\bibnamefont{W\"olfle}},
  \bibinfo{journal}{Phys. Rev. B} \textbf{\bibinfo{volume}{53}},
  \bibinfo{pages}{1850} (\bibinfo{year}{1996}).

\bibitem[{\citenamefont{Friedel}(1952)}]{Friedel1952}
\bibinfo{author}{\bibfnamefont{J.}~\bibnamefont{Friedel}},
  \bibinfo{journal}{Phil. Mag.} \textbf{\bibinfo{volume}{43}},
  \bibinfo{pages}{153} (\bibinfo{year}{1952}).

\bibitem[{\citenamefont{Langreth}(1966)}]{PhysRev.150.516}
\bibinfo{author}{\bibfnamefont{D.~C.} \bibnamefont{Langreth}},
  \bibinfo{journal}{Phys. Rev.} \textbf{\bibinfo{volume}{150}},
  \bibinfo{pages}{516} (\bibinfo{year}{1966}).

\end{thebibliography}

\end{document}